\documentclass{PoS}

\title{TopFitter: Fitting top-quark Wilson Coefficients to Run II data \\[-0.11mm]}

\ShortTitle{TopFitter}

\author{Stephen Brown\\
        SUPA, School of Physics and Astronomy, University of Glasgow, Glasgow, G12 8QQ, UK\\
        E-mail: \email{s.brown.7@research.gla.ac.uk}}

\author{Andy Buckley\\
        SUPA, School of Physics and Astronomy, University of Glasgow, Glasgow, G12 8QQ, UK\\
        E-mail: \email{a.g.buckley@gmail.com}}

\author{Christoph Englert\\
        SUPA, School of Physics and Astronomy, University of Glasgow, Glasgow, G12 8QQ, UK\\
        E-mail: \email{christoph.peter.englert@cern.ch}}

\author{James Ferrando\\
        DESY, D-22607 Hamburg, Germany\\
        E-mail: \email{james.ferrando@desy.de}}

\author{Peter Galler\\
        SUPA, School of Physics and Astronomy, University of Glasgow, Glasgow, G12 8QQ, UK\\
        E-mail: \email{Peter.Galler@glasgow.ac.uk}}

\author{\speaker{David J Miller}\\
        SUPA, School of Physics and Astronomy, University of Glasgow, Glasgow, G12 8QQ, UK\\
        E-mail: \email{david.j.miller@glasgow.ac.uk}}

\author{Liam Moore\\
        Centre for Cosmology, Particle Physics and Phenomenology (CP3), Universit\'{e} catholique de Louvain, B-1348 Louvain-la-Neuve, Belgium\\
        E-mail: \email{l.moore@cern.ch}}

\author{Michael Russell\\
        Institut f\"ur Theoretische Physik, Universit\"{a}t Heidelberg, Germany\\
        E-mail: \email{russell@thphys.uni-heidelberg.de}}

\author{Chris White\\
        Centre for Research in String Theory, School of Physics and Astronomy, Queen Mary University of London, 327 Mile End Road, London E1 4NS, UK\\
        E-mail: \email{christopher.white@qmul.ac.uk}}

\author{Neil Warrack\\
        SUPA, School of Physics and Astronomy, University of Glasgow, Glasgow, G12 8QQ, UK\\
        E-mail: \email{n.warrack.1@research.gla.ac.uk}}

\abstract{We describe the latest TopFitter analysis, which uses top quark observables to fit the Wilson Coefficients of the SM augmented with dimension-6 operators. In particular, we discuss the inclusion of new LHC Run II data, and the implementation of particle-level observables.}

\FullConference{ICHEP 2018, XXXIX International Conference on High Energy Physics, July 4-11, 2018, COEX, Seoul}

\begin{document}

\section{Standard Model Effective Field Theory}
The LHC has been successful, not only in discovering the Higgs boson, but also in extending the exclusion limits on new physics such as supersymmetry and technicolour. However, as thresholds for direct production of  exotic particles move beyond the LHC's energy, we must exploit additional methods in our search for new physics. Effective Field Theory (EFT) provides a model-independent framework to explore physics at a scale $\Lambda$ significantly above the LHC energy.

The Standard Model (SM) EFT is an extension of the SM to include higher dimensional operators (of dimension $D>4$) constructed from SM fields. The SM EFT Lagrangian becomes,
\begin{equation}
{\cal L}_{\rm SM~EFT}  = {\cal L}_{\rm SM} + \frac{C^{(5)}_1}{\Lambda} {\cal O}^{(5)}_1 + \frac{1}{\Lambda^2} \sum_i C_i^{(6)} {\cal O}^{(6)}_i + \ldots,
\end{equation}
where ${\cal O}_i^{(D)}$ denote operators of dimension $D$, $C_i^{(D)}$ are accompanying {\em Wilson coefficients}, and the ellipsis represent operators of higher dimension suppressed by increasing powers of $1/\Lambda$. {\em All} theories of new physics at a high scale $\Lambda$ that couples to SM particles, must manifest in this way at LHC energies, so setting limits on the Wilson coefficients provides extremely general exclusions. In this talk, I will report on our progress towards constraining the coefficients of operators involving top-quarks using LHC and Tevatron data. We collect top-quark observables sensitive to potential new physics, and ask how these observables may be affected by the additional operators of the SM EFT. We then fit to the data to extract bounds on the corresponding Wilson coefficients.

The operator ${\cal O}_1^{(5)}$ is interesting to neutrino physics but not relevant to top-quarks; here we will be interested in the dimension-six operators only. Assuming minimal flavour violation and Baryon number conservation, there are 59 independent dimension-six operators in the SM EFT~\cite{Grzadkowski:2010es}, but only 16 of these are relevant for top-quark physics. Some are not constrained by data or only constrained in particular combinations. Indeed, if we do not include $Wt$ associated production, we have 6 operators constrained by top-pair, and 3 operators constrained by single-top and top-decays.

In section~\ref{sec:run1}, I will briefly review our previous work using the Tevatron and LHC Run I data~\cite{Buckley:2015nca,Buckley:2015lku}. In section~\ref{sec:improvements} I will outline the improvements we intend to make for Run II, and in section~\ref{sec:trial} I will describe a trial study using the data from a single ATLAS paper.

\section{TopFitter and Run I \label{sec:run1}}
Our first TopFitter analysis~\cite{Buckley:2015nca,Buckley:2015lku} used results from the ATLAS and CMS $7$ and $8\,$TeV datasets as well as the Tevatron. We also used a single Run II analysis of the top-quark pair-production cross-section at $13\,$TeV from CMS~\cite{Khachatryan:2015uqb}. In total, we included 227 different measurements, most of which (195) were top-pair production, but also considered single top and associated production. We didn’t include $Wt$ associated production, due to the difficulties of disentangling it from top-pair production at next-to-leading order. The majority of these observables are differential distributions and all of them are corrected back to the parton level (including only direct decay products of the top). We incorporate the analyses systematic and statistical uncertainties, adding them in quadrature, and include correlations between measurements were available. However, we note that full correlations have not been made available for many of of the analyses, potentially making our exclusions stronger than they should be. For a list of the measurements used, see Table~1 of Ref.~\cite{Buckley:2015lku}

To make a comparison of this data with the SM EFT, we must provide theory simulations in the space of Wilson coefficients. We first implemented the SM EFT Lagrangian in FeynRules~\cite{Alloul:2013bka}, and feed the result into MadGraph~5~\cite{Alwall:2014hca} to provide Leading Order (LO) parton-level observables. These are upgraded to Next-to-Leading Order (NLO) by applying bin-by-bin K-factors as calculated by MCFM~\cite{Campbell:2010ff}, and we also include Next-to-Next-to-Leading Order (NNLO) contributions where appropriate. Theoretical uncertainties are estimated by varying renormalisation and factorisation scales, as well as using a variety of parton distribution functions. This is done over the entire Wilson coefficient space $\mathbf{C}= \{C_i\}$, logarithmically sampled.

We then construct a polynomial parameterising function $f_b(\{C_i\})$ for each observable bin $b$, which fits the sampled points with least-squares-optimal precision. This allows us to interpolate to any Wilson coefficiant choice and we can construct a $\chi^2$ distribution according to,
\begin{equation}
  \chi^2({\mathbf{C}})=\sum_{\cal O} \sum_{i,j}
  \frac{
  \left(f_i(\mathbf{C})-E_i\right)
  \rho^{-1}_{i,j}
  \left(f_j(\mathbf{C})-E_j\right)
  }{\sigma_i \sigma_j},
\end{equation}
where $E_i$ are the experimental data values, and $\sigma_i$ includes theoretical and experimental uncertinties added in quadrature. This interpolation between discrete Wilson coefficient choices, and the fitting is done within the PROFESSOR framework~\cite{Buckley:2009bj}.

We used this to constrain 9 top-quark operators (6 operators constrained by top-pair, and 3 constrained by single-top and decays) that are linear combinations of those in Ref.~\cite{Grzadkowski:2010es}. We presented both individual constraints, where the oparators are turned on one at a time, and also marginalised constraints, were they are all turned on simultaneously. Marginalised constraints are weaker than individual ones since contributions to an observable from one operator may be cancelled by those from another. These results can be seen in Figure 6 (left) of Ref.~\cite{Buckley:2015lku}. We found a good quality of fit and no significant tension with the SM.

%%%%%%%%%%%%%%%%%%%%%%%%%%%%%%%%%%%%%%%%%%
\begin{figure}[!h]
\begin{center}
\includegraphics[width=0.5\textwidth]{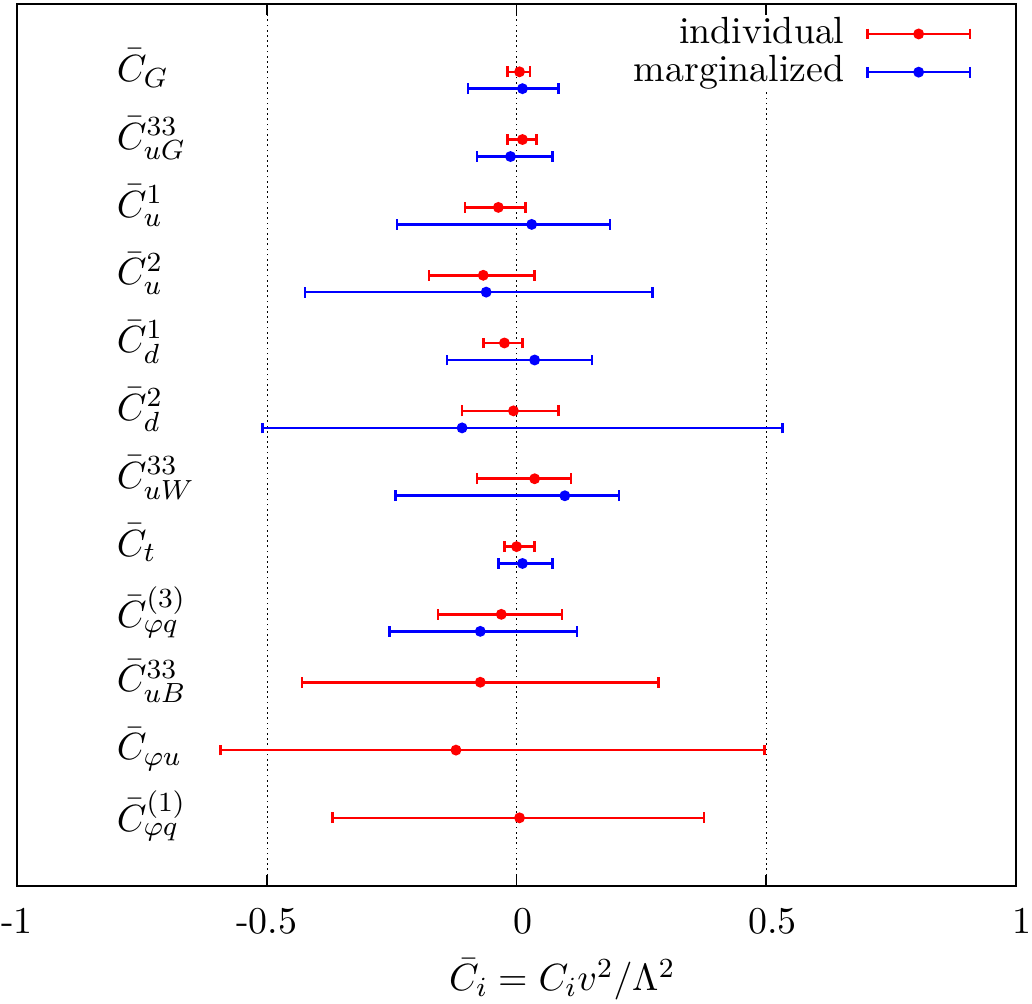}
\caption{95\% confidence constraints on Wilson coefficients from Run I and Tevatron data. This figure is taken from Ref.~\cite{Buckley:2015lku}, where one can also find a description of these Wilson coefficients combinations.}
\label{fig:run1_constraints}
\end{center}
\end{figure}
%%%%%%%%%%%%%%%%%%%%%%%%%%%%%%%%%%%%%%%%%%

\section{Improvements for Run II \label{sec:improvements}}
In extending TopFitter to Run II we would like to improve on our analysis. In particular, most new measurements are presented at the particle level, so we should go beyond parton-level and include parton showering in our simulaton. We also want to include all possible observables, including $Wt$, which means sampling over the entire top-quark Wilson coefficient space. However, this ambitious goal is computationally impractical using the method we used for the Run I analysis.

To overcome this we have {\em linearised} our simulation in the Wilson coefficients. Previously we calculated $| {\cal M}_{\rm SM} + {\cal M}_{\rm HDO}|^2$ for every sampled point in our coefficient space, where ${\cal M}_{\rm SM}$ and ${\cal M}_{\rm HDO}$ represent the matrix elements containing the SM and Higher Dimensional Operators (HDO) respectively. Instead we expand this square to give
$|{\cal M}_{\rm SM}|^2 + 2 {\rm Re} \left( {\cal M}_{\rm SM}^\dagger {\cal M}_{\rm HDO} \right) + |{\cal M}_{\rm HDO}|^2$.
The first term is just the SM and can be calculated once for each observable; the second term is {\em linear} in each of the Wilson coefficients; while the third term is higher order in $1/\Lambda$ and may (formally) be neglected. The interference term of each operator with the SM can be calculated once for each observable, and scaled to provide a theoretical prediction for any required Wilson coefficiant chocie without need for interpolation. This makes the generation of theory predictions and the subsequent fitting much faster, and a particle level analysis over the entire space becomes feasible.

We note that neglecting $|{\cal M}_{\rm HDO}|^2$ could be problematic. Although this is formally of higher order, if the interference of the operator with the SM is small or zero (e.g.\ due to colour or helicity conservation) then this neglected term could provide the leading contribution of new physics. Furthermore, it may provide the leading contribution in some regions of phase space. However, one could avoid these issues by also calculating $|{\cal M}_{\rm HDO}|^2$ for each operator (or pair of operators); since we know how this scales with the Wilson coefficient, these also only need to be caclulated once.

\begin{figure}[!h]
%\begin{center}
\includegraphics[width=0.5\textwidth]{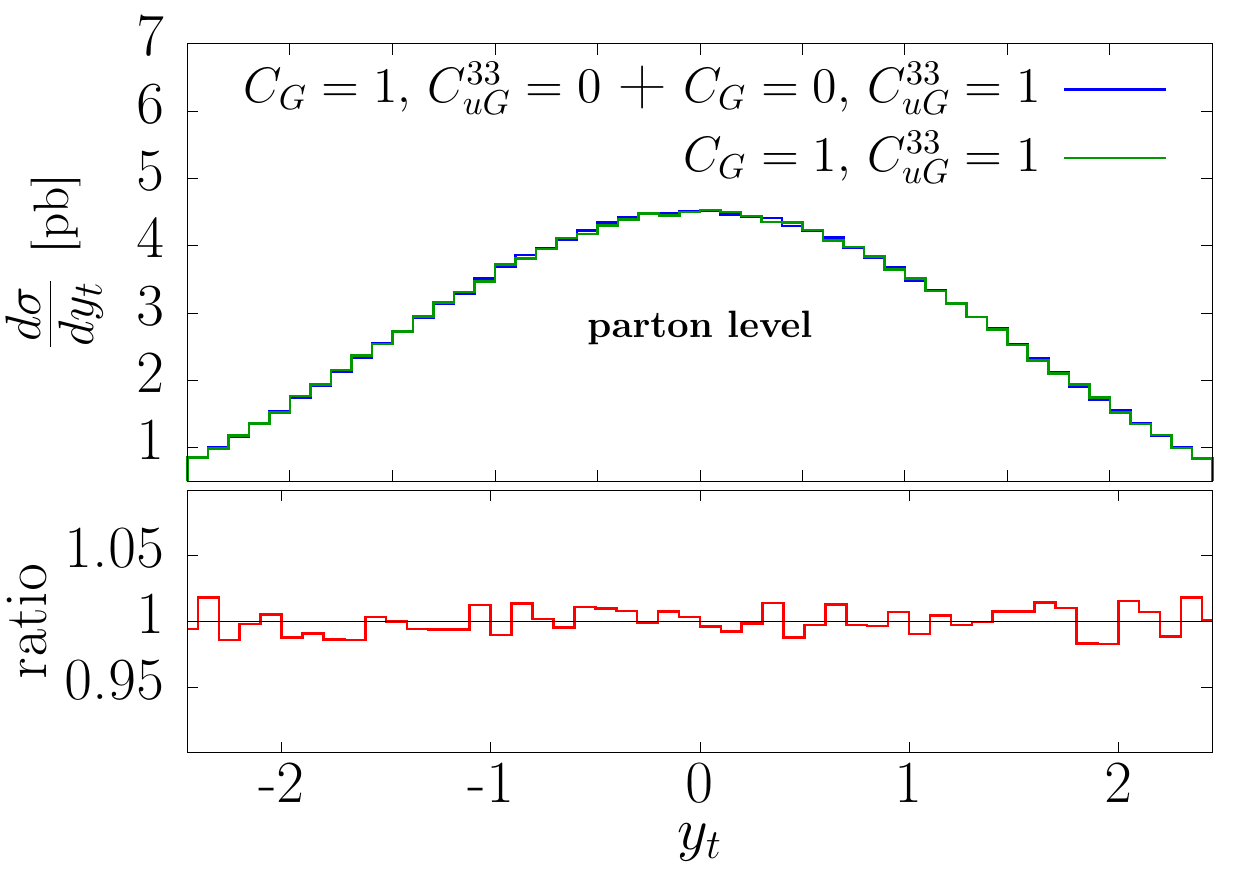}
\includegraphics[width=0.5\textwidth]{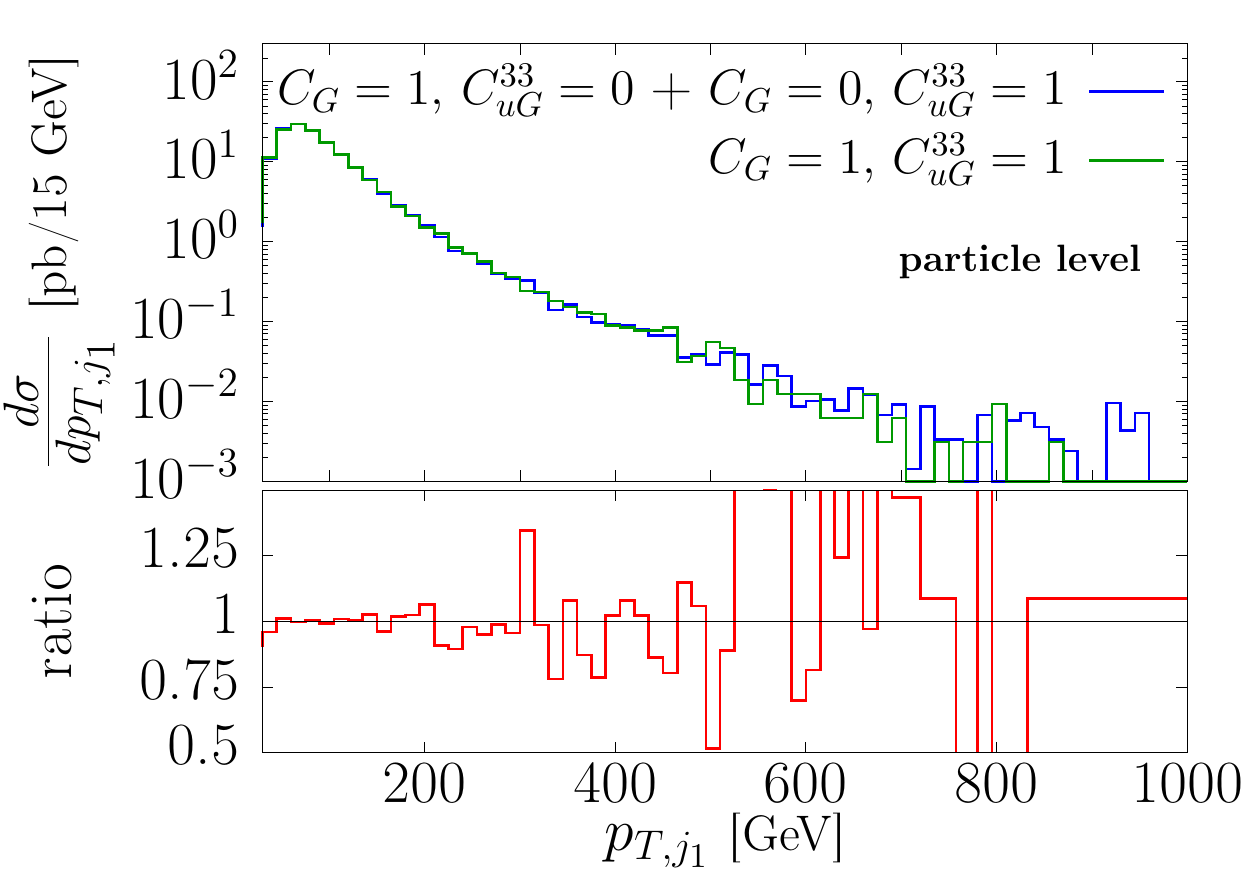}
\caption{A test of the linearisation in the Wilson coefficents, at parton level (left) and particle level (right).}
\label{fig:lin_test}
%\end{center}
\end{figure}

We have tested this linearisation by generating $13\,$TeV NLO parton-level events using MadGraph 5 and showering them with Pythia 8. The tops are decayed using MadSpin~\cite{Artoisenet:2012st}, while $t \bar t$ and $t \bar t\,+\,$jet are merged and matched with the parton shower using the procedure of Ref.~\cite{Frederix:2012ps}, and include the jet matching scale in our uncertainties. Example distributions can be seen in Figure~\ref{fig:lin_test} comparing the previously used method with the new linearised method. On the left we see the rapidity calculated at the parton level including two non-zero Wilson coefficients simultaneously (blue) compared with taking one Wilson coefficent non-zero at a time and adding the interferences together (green). We see an extremely good match. This agreement is also seen for the $p_T$ of the leading jet on the right, now tested at the particle level.

%%%%%%%%%%%%%%%%%%%%%%%%%%%%%%%%%%%%%%%%%%
%%%%%%%%%%%%%%%%%%%%%%%%%%%%%%%%%%%%%%%%%%

%%%%%%%%%%%%%%%%%%%%%%%%%%%%%%%%%%%%%%%%%%
% \begin{figure}[!h]
% %\begin{center}
% \includegraphics[width=0.5\textwidth]{pT_leading_top_Cu1.pdf}
% \caption{...}
% \label{fig:lin_test_parton}
% %\end{center}
% \end{figure}
%%%%%%%%%%%%%%%%%%%%%%%%%%%%%%%%%%%%%%%%%%

\section{A Run II Trial using ATLAS Data \label{sec:trial}}
\begin{figure}[!h]
%\begin{center}
\includegraphics[width=0.5\textwidth]{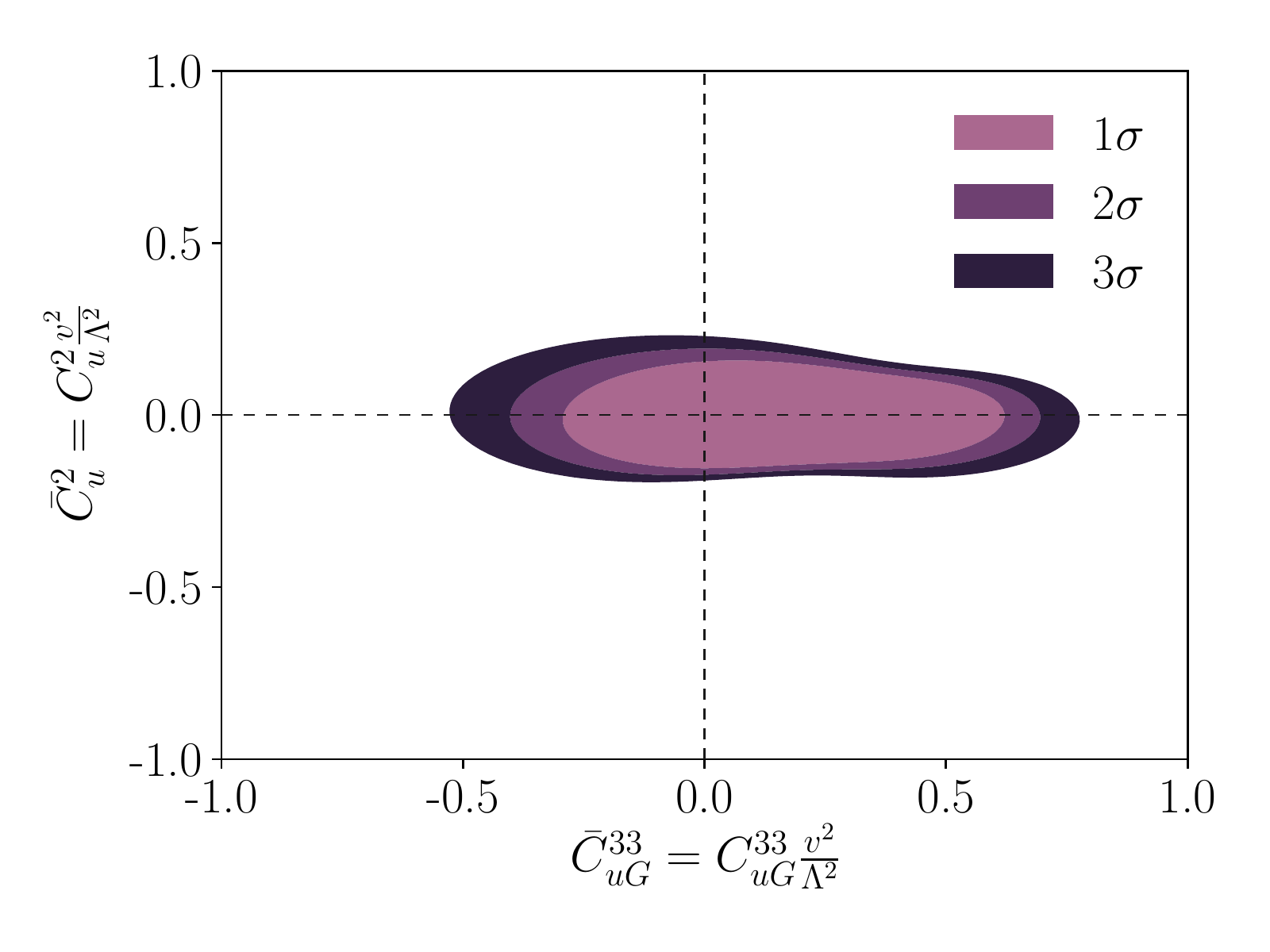}
\includegraphics[width=0.5\textwidth]{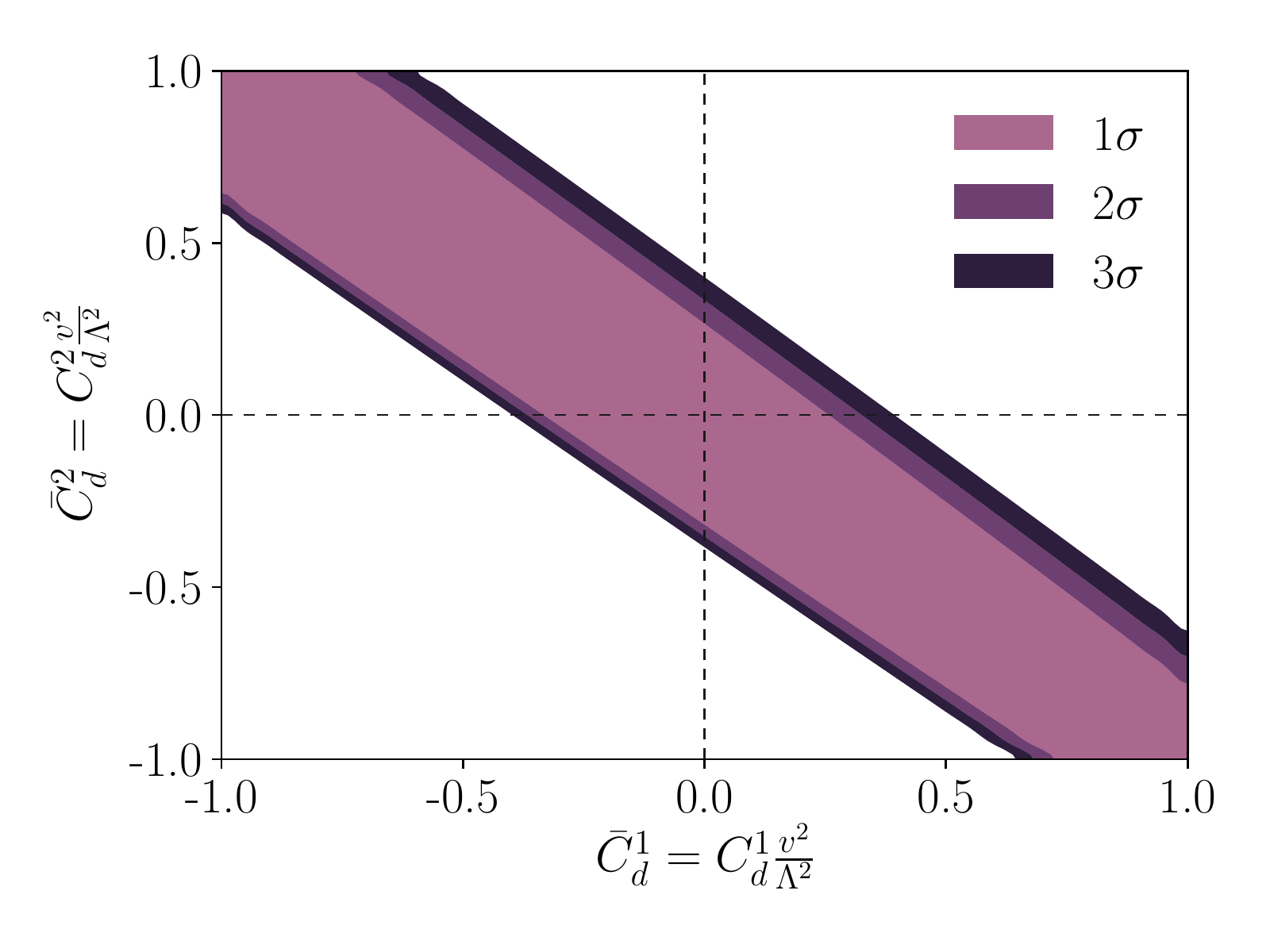}
\caption{Constraint contours at $1\, \sigma$ (red),
$2\, \sigma$ (green) and $3\, \sigma$ (blue) on pairs of Wilson coefficients from our test analysis on a single ATLAS paper~\cite{Aaboud:2018eqg}.}
\label{fig:contours}
%\end{center}
\end{figure}

\begin{figure}[!h]
\begin{center}
\includegraphics[width=0.5\textwidth]{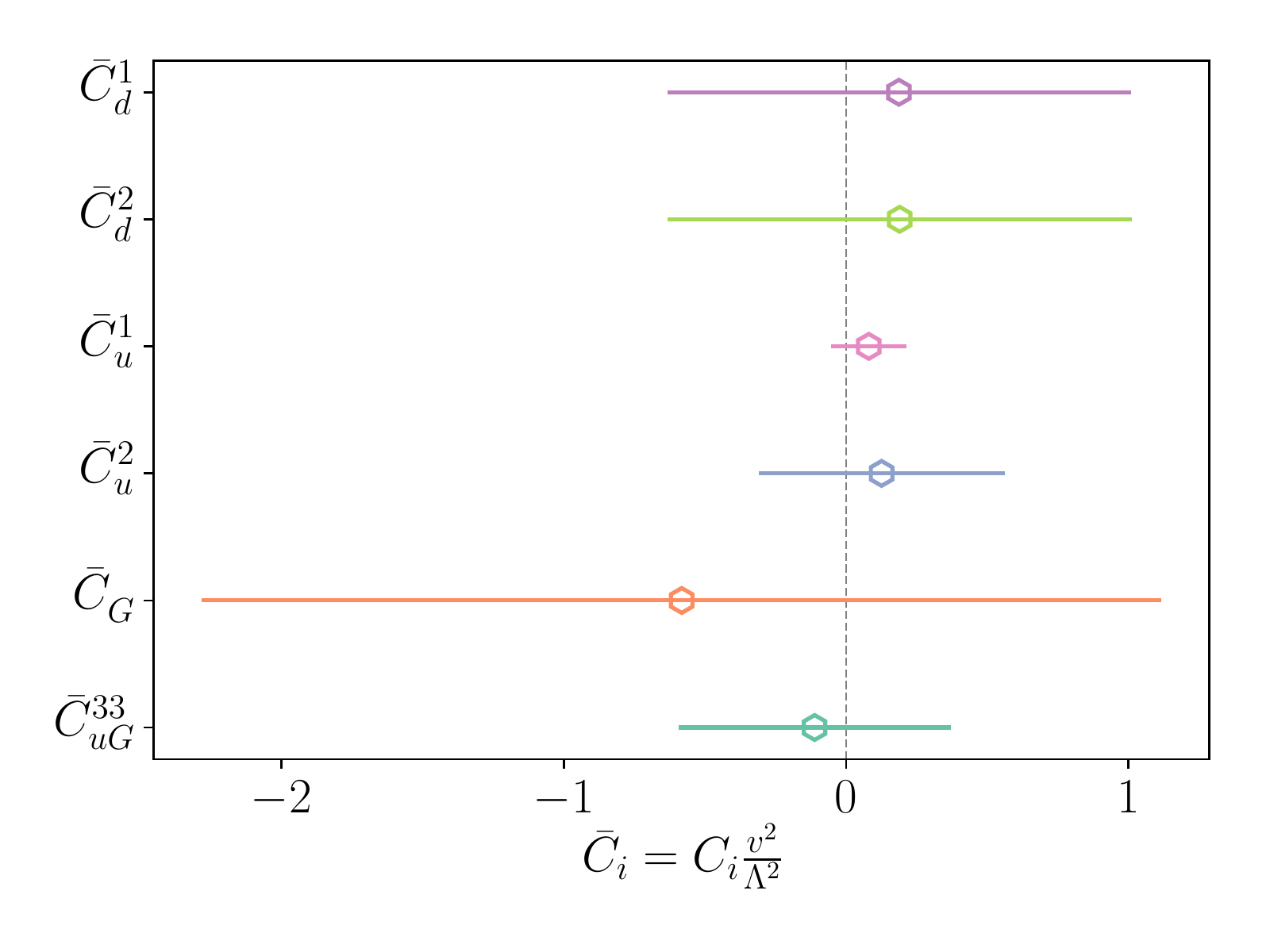}
\caption{95\% confidence constraints on individual Wilson coefficients from our test analysis on a single ATLAS paper~\cite{Aaboud:2018eqg}. These constraints are {\em not} marginalised.}
\label{fig:run2_constraints}
\end{center}
\end{figure}

To test our new methodology further we have performed a fit of the SM EFT to the data presented in the ATLAS analysis of Ref.~\cite{Aaboud:2018eqg}, which presents 13 differential observables in top-pair production at $13\,$TeV. We chose this particular analysis because it is fully available on HEPData, has RIVET~\cite{Buckley:2010ar} analyses for its observables, and has full correlations between observables and between bins\footnote{Unfortunately, due to a technical issue we were unable to include these correlations in the results shown here, but do intend to include them in the future.}. We used the same methodology described in Section~\ref{sec:improvements}, and present very preliminary example results in Figures~\ref{fig:contours} and \ref{fig:run2_constraints}. In Figure~\ref{fig:contours} we see constraints on two operators simultaneously, where the red, green and blue extents represent $1$, $2$ and $3\,\sigma$ exclusions respectively. We can see, in the right-hand plot, that some directions in Wilson coefficient space remain unconstrained by these particular measurements. In Figure~\ref{fig:run2_constraints} we see all the constraints from this new data on six of the the Wilson coefficents we previously saw in Figure~\ref{fig:run1_constraints}. We note that these are only non-marginalised constraints here, and are understandably weaker than those of the Run I analysis since there is much less data. Nevertheless, this demonstrates a good test of our methodology.

%%%%%%%%%%%%%%%%%%%%%%%%%%%%%%%%%%%%%%%%%%
%%%%%%%%%%%%%%%%%%%%%%%%%%%%%%%%%%%%%%%%%%

%%%%%%%%%%%%%%%%%%%%%%%%%%%%%%%%%%%%%%%%%%
%%%%%%%%%%%%%%%%%%%%%%%%%%%%%%%%%%%%%%%%%%

\section{Conclusions}
We have presented here the first preliminary results of TopFitter for Run II LHC data. Our analysis has been considerably improved in comparison to that for Run I, now allowing the inclusion of particle-level data. To do this we have adopted a method of linearisation in the Wilson coefficents, which we have comprehensively tested. We have produced preliminary constraints on the Wilson coefficients from a single ATLAS analysis. This analysis should be improved by including the full correlations and marginalising over the operators. Then we will embark on a full analysis of Run II data, and produce general constraints of new physics in the top sector.

\end{document}